# Chargeable photoconductivity in Van der Waals heterojunctions

Yucheng Jiang[1,5,*], Anpeng He[1,5], Yu Chen[1], Guozhen Liu[1], Hao Lu[1], Run Zhao[1], Mingshen Long[2], Ju Gao[3], Quanying Wu[1], Xiaotian Ge[4], Jiqiang Ning[4] and Weida Hu[2]

[1]Jiangsu Key Laboratory of Micro and Nano Heat Fluid Flow Technology and Energy Application, School of Mathematics and Physics, Suzhou University of Science and Technology, Suzhou, Jiangsu 215009, PR China

[2]State Key Laboratory of Infrared Physics, Shanghai Institute of Technical Physics, Chinese Academy of Sciences, 500 Yu Tian Road, Shanghai 200083, China.

[3]School for Optoelectronic Engineering, Zaozhuang University, Shandong 277160, China

[4]Vacuum Interconnected Nanotech Workstation, Suzhou Institute of Nano-tech and Nano-Bionics (SINANO), Suzhou, Jiangsu 215009, PR China

[5]These authors contributed equally: Yucheng Jiang, Anpeng He.

*e-mail: jyc@usts.edu.cn

*Abstract*

Van der Waals (vdW) heterojunctions, based on two-dimensional (2D) materials, show great potential for the development of eco-friendly and high-efficiency nano-devices. Considerable research has been performed and has reported valuable applications of photovoltaic cells, photodetectors, etc. However, simultaneous energy conversion and storage in a single device has not been achieved. Here, we demonstrate a simple strategy to construct a vdW *p-n* junction between a $WSe_2$ layer and quasi-2D electron gas. After once optical illumination, the device stores the light-generated electrons and holes for up to seven days, and then releases a very large photocurrent of 2.9 mA with bias voltage applied in darkness; this is referred to as chargeable photoconductivity (CPC), which completely differs from any previously observed photoelectric phenomenon. In normal photoconductivity, the recombination of electron-hole pairs takes place at the end of their lifetime, causing a release of heat; in contrast, infinite-lifetime photocarriers can be generated in CPC devices without a thermal loss. The photoelectric conversion and storage are completely self-excited during the charging process. The ratio between currents in full- and empty-energy states below the critical temperature reaches as high as $10^9$, with an external quantum efficiency of 4410000% during optical charging. A theoretical model developed to explain the mechanism of this effect is in good agreement with the experimental data. This work paves a path towards storage-type photoconductors and high-efficiency entropy-decreasing devices.

**Introduction**

Photoelectric conversions, such as photoconductive and photovoltaic effects, often occur in *p-n* junction systems, where incident photons create electron-hole pairs (EHPs) [1,2]. For high-performance devices, it is necessary to achieve high photoelectric and EHP separations [3]. However, rapid recombination of EHPs tends to shorten the lifetime of photocarriers, therefore, limiting external quantum efficiency (EQE) [4,5]. Many strategies have been proposed to obtain efficient photoelectric devices by searching for novel *p*- and *n*-type materials or designing a functional interface structure [6-9]. Basic methods of realizing high EQE involve the improvement of the trapping or absorption of photons, e.g., surface plasmon excitation and element doping; few studies have focused on extending the lifetime of photocarriers [10-15]. It is yet unknown whether infinite-lifetime photocarriers can exist stably in a photoelectric system.

Generally, the junction trapping plays an important role in enhancing the lifetime of photocarriers and controlling the separations of EHPs. Recent advances in Van der Waals (vdW) heterojunctions have created new potential for the development of high-performance photoelectric devices [16-22]. Based on two-dimensional (2D) layered materials, the vdW heterostructures give rise to attractive possibilities for manipulating the generation, recombination and transport of photocarriers in atomic interfaces [16,20,21]. Several notable photoelectric properties have been reported by facilitating the design of atomically thin devices, such as photocatalysis [19], photoconductivity (PC) [22,23], photovoltaic effect [24,25], and electroluminescence [26]. A strong advantage is that the layer-dependent band structure allows the physical properties of the *p-n* junction to be controlled without creating extra disorder [27]. Beyond 2D/2D heterostructures, mixed-dimensional vdW junctions have also attracted some interest, which aims to combine the advantages of 2D and conventional materials [16,18,22]. Until now, most of vdW junctions are mainly structured by *n*- or *p*-type 2D layered materials [20]. Another 2D system, 2D electron gas, has been extensively studied because of its abundant physical properties [28-30], but no effort has been made to construct *p-n* junctions based on it.

Here, we develop a practical method to establish a lateral vdW *p-n* junction between few-layered $WSe_2$ and quasi-2D electron gas (Q2DEG) on $SrTiO_3$ (STO). The chargeable photoconductivity (CPC) effect is observed for the first time, completely differing from any previously observed photoelectric phenomenon. The device can not only create photocarriers under optical illumination, but also store them, indicating the achievement of simultaneous energy conversion and storage. Specifically, photocarriers are stored in the space charge region (SCR) without recombination. After at least seven days, a large photocurrent of 2.9 mA is released with bias applied in darkness. During this process, EHPs recombine and the device returns to the insulating state. The ratio of currents in full- and empty-photocarrier states reaches as high as $10^9$ with high EQE of 4410000%. Based on the existence of infinite-lifetime photocarriers, we propose a theoretical model to explain the mechanism of this remarkable effect, which is in good agreement with experimental data.

**Results**

A schematic of the $WSe_2$/Q2DEG heterostructure is shown in Fig. 1a. During the fabrication, one half of few-layered WSe$_2$ is coated with photoresist, and the other half is irradiated by an Ar$^+$ ion beam (see Supplementary Fig. 1). A thin layer of oxygen vacancies will be induced by Ar$^+$ etching on the STO surface. From previous research, oxygen vacancies tend to cause a high density of Q2DEG [31,32]. In this case, a vdW contact can be achieved between the edges of few-layered WSe$_2$ and Q2DEG. Despite indirect contact, the electric transport is still realized by the electron tunneling. Due to the *p*-type conduction of intrinsic WSe$_2$, a lateral *p-n* heterojunction is formed in the interface of few-layered WSe$_2$ and Q2DEG, exhibiting good rectifying behavior (see Supplementary Fig. 2a). In Fig. 1b, an atomic-force microscopy (AFM) image shows the surface morphology of the device with the WSe$_2$ flake of 39.5 nm thickness. The inset is a photographic image of the selected region. It is note that the shape of WSe$_2$ is stamped on the etched region, although there is no WSe$_2$ left after the Ar+ ion bombardment. Because of the simultaneous etching of WSe$_2$-covered and exposed STO regions, the exposed STO region has been etched for some moments when the WSe$_2$ is completely etched, causing that the WSe$_2$-covered region is higher than the other region.

With the circuit cut off, the device is irradiated by a 405 nm, 16 mW/cm$^2$ laser for 3 s at 30 K, as shown in Fig. 1a. The light is then turned off, and the device is kept in darkness for 2 min or seven days, as shown in Fig. 1c. Afterwards, a very large photocurrent of about 3.1 mA is released with bias voltage increasing to 5 V in darkness, as shown in Fig. 1d. To demonstrate the reliability of this result, the whole measurement process has been provided in Supplementary Movie. Note that the waiting process shown in Fig. 1c must be performed in a circuit cut-off state; otherwise, photocarriers will be slowly released, even at zero voltage (see Supplementary Fig. 2b). It seems that the device is self-excited to store photocarriers in the *p-n* interface under optical illumination, which is highly similar to the charging process of a battery. After optical charging, the stored photocarriers can flow across the junction under bias voltage, even with the light removed. A complete current-voltage (I-V) loop contains the forward (voltage increase) and backward (voltage decrease) curves. The forward curve shows a releasing process (RLP) of the photocarriers, and the backward one implies a recovery process (RCP) towards the vacancy state of the carriers. The inset in Fig. 1d demonstrates that the maximum current under optical illumination is $10^9$ times higher than that in darkness. Within one I-V loop, the current ratio of RLP and RCP can be as high as $10^4$ at 3 V, which demonstrates that over 99.9% of stored photocarriers recombine under bias voltage. In this work, we focus on the CPC effect at positive bias voltages, although the device can also discharge photocarriers at negative bias voltages, as shown in Supplementary Fig. 3a. It is observed that the increase of bias voltage can cause longer discharge time in Supplementary Fig. 3b. A simple explanation is that the larger voltage will drive more storage charge to participate in the recombination, which inevitably costs more time. The negative bias shows even higher efficiency of discharge than the positive bias (see Supplementary Note 1). Once photocarriers are consumed in this situation, no significant photocurrent will be obtained at positive voltages. Moreover, the cycle of charge and discharge is highly repeatable. The device enables to maintain the same level of photocurrent even after 1000 cycles (see Supplementary Fig. 4a).

Fig. 2a shows I-V curves in RLP with different bias voltage increment rates ($v_r$), offering additional evidence of storage behavior. It is found that current increases significantly with the increase of $v_r$, which differs from the normal photoconductivity (NPC) such as instantaneous and persistent photoconductivity effects. For this CPC effect, larger $v_r$ means a stronger tendency to recombine EHPs and shorter carrier release time. This inevitably leads to an increase in current, due to the fixed amount of stored charges with full charging. The $v_r$ dependence of current is a key feature that can be used to distinguish CPC from conventional photoelectric responses in experiments.

I-V loops are repeated five times after a single optical charge, as shown in Fig. 2b. Although over 99% of stored electrons and holes can be consumed in the first loop, residual charges are still trapped in the intrinsic space charge region (ISCR) to prevent their return to an intrinsic insulating state. Besides, the photo-switching characteristic shows fast photoelectric responses, indicating a different nature from persistent photoconductivity despite their similar ability to record optical information (see Supplementary Fig. 4b and Supplementary Notes 2 and 3). The response time is less than 30 ms by investigating the photocurrents with on/off light (see Supplementary Fig. 5a). The inset in Fig. 2b shows the I-V curve with a light on, compared with that in darkness after full charge. Ordinarily, for the NPC effect, photoelectric materials should exhibit the largest photocurrent under the continuous irradiation of light. However, a different phenomenon has been observed in the WSe$_2$/Q2DEG heterostructure, by which the discharge current measured in darkness is much larger than that under continuous optical illumination. A reasonable explanation may be offered on the basis of the infinite lifetime of the photocarriers. Without EHP recombination, the photocarriers will gradually accumulate to a high density, thus causing a large photocurrent in darkness. By contrast, with bias voltage and optical illumination applied together, the light-produced electrons and holes tend to be recombined promptly without accumulation. Another possible mechanism is that newly-produced light-produced electrons and holes tend to diffuse to positive and negative ISCR, respectively. The as-formed diffusion current flows in the opposite direction of the drift current under forward bias voltage, which causes the total current to decrease.

At low temperature, the device cannot return to electric conduction, with the forward bias increasing to values as large as 25 V. This phenomenon may be attributed to a large built-in potential over 20 V and a long ISCR with the width of 18 μm. in the p-n junction. This type of ISCR seems to act as a container to store the photocarriers. In order to clarify the mechanism of photocarrier storage, Fig. 2c shows I-V loops measured one by one in a series of voltage ranges from 0–4 V to 0–16 V after a single full charge. A small $v_r$ is used to prevent damage to the device due to a large current. Intriguingly, after the stored photocarriers are completely exhausted in the low voltage ranges, the device still releases a large photocurrent at higher voltages. High bias voltage may drive infinite-lifetime photocarriers, stored in ISCR, to cross built-in barriers, thus contributing to the electrical transport. Here, the maximum photocurrents of all loops are plotted to indicate the distribution of photocarrier density at different voltage levels. It is observed that the photocarrier density increases with increase in the voltage level, and reaches saturation above 10 V. Moreover, photocarriers at different voltage levels are simultaneously generated by incident light. This feature signifies the high charging efficiency and large storage capacity of photocarriers.

Fig. 3a illuminates photocurrent and dark currents as a function of temperature at 4 V. With temperature dropping, the $WSe_2$/Q2DEG heterostructure gradually loses its electric conductivity, and even acts as an insulator below 160 K. By contrast, the photocurrent tends to increase sharply as the temperature decreases to 9 K. It is noted that there exists a critical temperature ($T_c \approx 80.3$ K) below which the photocurrent exhibits a rapid increase. For simplicity, we use $Q_{m-n}$ to express the charge capacity of photocarriers between the voltage levels of *m* and *n*. Due to the insulation after complete discharging, all of stored charges will become the carriers collected in the electric measurement. It is reasonable to estimate the number of stored charges by measuring current. So we have $Q_{m-n} = \int_{t=0}^{t=t_n} I(t)dt - \int_{t=0}^{t=t_m} I(t)dt$ , where $t_m$ and $t_n$ are the time taken to increase the bias voltage to *m* and *n* V from zero, respectively. Fig. 3b shows the temperature dependence of $Q_{0-25}$. It is found that the storage capacity of photocarriers increases rapidly as temperature decreases below 80.3 K, a temperature which happens to be the same as the $T_c$. In order to achieve the CPC effect, two requirements have to be met: electric insulation and infinite-lifetime photocarriers.

WSe$_2$/Q2DEG heterostructures become insulating below 160 K and produces infinite-lifetime photocarriers below T$_c$. T$_c$ may be a phase transition temperature at which the lifetime of photocarriers becomes infinity, and the device changes from an NPC state to CPC state (see supplementary Note 4). Here, the infinite lifetime mainly refers to the storage lifetime of photocarriers (see Supplementary Note 5 and supplementary Fig. 5b). In addition, the charge capacity of Q$_{0-5}$ shows a dependence on thickness. Several devices have been fabricated in the thickness range from 10 to 100 nm (see Supplementary Fig. 6). Among them, the 39-nm device shows the best photoelectric response, and is used throughout the experiments.

Fig. 3c shows I-V loops after full charging with different wavelengths of visible light. As expected, the short-wavelength (or high-energy) photons are more likely to transform into infinite-lifetime photocarriers in the junction; in other words, the CPC device can not only record the history of optical illumination, but also identify wavelength information. For long-wavelength (over 532 nm) visible light, the CPC effect is still significant, but is achieved only below the bias voltage of 21 V. It seems that the energy of long-wavelength is too low to activate the storage behavior of photocarriers at high voltage levels. In addition, the optical charging procedure has been investigated in Fig. 3d Despite having different optical powers, values of Q$_{0-4}$ can reach almost the same maximum value with sufficient exposure time, suggesting full optical charging. The charging time depends strongly on the optical power density. Under a 5 mW/cm$^2$ optical illumination, it takes less than 6 s for the device to be fully charged. As the optical power density decreases, more time is needed for full charging. No matter how small the optical power is, the device always harvests sporadic photons and accumulates plenty of photocarriers, finally releasing a large photocurrent under bias voltages. This ability of "many a little makes a mickle" distinguishes the CPC effect from other photo-electrical processes. Similar to an entropy decrease process, the WSe$_2$/Q2DEG heterostructure enables the scattered and disordered photons in an environment to be captured, turned into electrons and holes, and have the energy stored in the junction. Here, we use EQE to evaluate the charging ability of the device, which is defined as the ratio of the number of collected light-generated carriers to the number of incident photons. The self-excited transition procedure from photons to infinite-lifetime photocarriers exhibits high EQE, up to 4410000% (see Supplementary Fig. 7). Generally, EQE exceeding 100% may suggest multiple electron-hole pairs created by a single photon [3,33], but this single factor is not sufficient to explain such large EQE values, exceeding $10^6$%. A long lifetime may cause photocarriers to circulate thousands of times across the channels before their recombination [34]. In this case, EQE is usually determined by $\tau/t_r$, where $\tau$ is the lifetime of photocarriers and $t_r$ is the transmit time.

To clarify the mechanism of charge storage, it is necessary to investigate the generation and distribution of stored photocarriers in ISCR. Here, we use a micro laser beam (MLB) with 0.7-μm radius to illuminate different positions on the heterojunction. In general, the current size at different voltages can reflect the distribution of carriers at different depths of ISCR. In Fig. 4a, six positions are respectively exposed to MLB, which are labelled as A to F. The photograph is exhibited in Supplementary Fig. 8. In this situation, we first investigate the effect of light illumination position on the charging process. The charging time is 40 s with the light power of 5 μW. A natural idea is that only the position, where the light reaches, can store the photocarriers. However, the experimental results disagree with this expectation as shown in Fig. 4b. Despite the different positions of light illumination, all of the I-V loops show almost the same turn-on voltages and current trends, which implies that the position of light illumination is not the location of photocarrier storage. It seems that stored photocarriers can be redistributed in ISCR to achieve lower energy status. Another important point is that no discharge current is observed at the position of A, which indicates that the region of Q2DEG on STO cannot store the photocarriers. Thus, the persistent optically gating effect in STO cannot be a factor to cause the CPC effect[35-37].

The redistribution of stored photocarriers in the charging makes it impossible to obtain the information about the positions of photocarrier generation from Fig. 4b. To clarify the relevant mechanism, continuous MLB is used to irradiate the position from B to F. Fig. 4c shows the I-V characteristics with MLB moving to the interface from far to near. It is worth noting that the voltage for the measurements should decrease from 20 V to 0 V, which eliminates the effect of stored photocarriers on the current. In this situation, the turn-on voltage tends to increase with the shortening of the distance with MLB and interface. There is a corresponding relationship between the turn-on voltage and the distribution of carriers. It is evident that the position of light illumination is exactly where photogenerated carriers are generated. The difference between the generation and storage positions of photocarriers implies that the PGHs can flow spontaneously to lower energy positions no matter where to create them. Fig. 4d exhibits the turn-on voltages as a function of W, demonstrating that the PGHs prefer to fill the region closer to the interface in ISCR (see Supplementary Fig. 9a). The injection of non-equilibrium carriers (PGHs) leads to the increase of local carrier density. The diffusion effect overcomes the impediment of built-in electric field, so PGHs move to the interface driven by diffusion (see Supplementary Note 6). Also, we can obtain the spatial information of ISCR by investigating the generation positions of photocarriers. Fig. 4 shows that the $WSe_2$ flake between the Au electrode and interface is in ISCR, from which the width of ISCR (about 18 μm) can be estimated. Based on the above understanding, we can figure out the processes of charge and discharge in Supplementary Notes 7 and 8 and Supplementary Fig. 9 b and c.

**Discussion**

On basis of the experimental results, two basic assumptions are suggested to build a theoretical model: (a) the lifetime of photocarriers is infinite; (b) the ISCR has a very large built-in potential ($V_{bi}$) and width, which can be regarded as a container to store photocarriers. As shown in Fig. 5a, the photo-generated holes (PGHs) will fill the ISCR so as to form a photocarrier-induced pseudo space charge region (PPSCR). According to the discussion in Supplementary Note 8, the three regions determine the electric transport of the device together, which are n region, p region and the PGH storage region (PSR). By the classical theory of *p-n* junctions, the photo-generated carriers per volume is written as $n_{ph} = G\tau$, where *G* is the generation rate of electron-hole pairs, and $\tau$ is the lifetime of photo-generated carriers. However, for infinite-lifetime photocarriers in the CPC effect, the above equation needs modification and is expressed as $n_{ph} = Gt_e$, where $t_e$ can be regarded as charge time. Since GPHs tend to move towards the interface, $n_{ph}$ should be the GPH density in PSR. The vdW interface and space charges tend to hold the photocarrier density steady and prevent the recombination of EHPs.

As per classical theory, there should exist a photocarrier-induced pseudo built-in potential ($V_{bi}^{ph}$) in PPSCR. Considering $V_{bi}^{ph} = V_{pn} + V_{pp}$ in supplementary Note 8, I-V characteristics depend mainly on $V_{bi}^{Ph}$. Moreover, it is evident that $V_{bi}^{Ph}$ can be increased with the release of photocarriers. Current flow causes the recombination of electron-hole pairs and consumes the PGHs in PPSCR. Since $V_{bi}^{ph}$ is directly determined by the distribution of photocarriers stored in ISCR, the consumption of PGHs will lead to an increase of $V_{bi}^{ph}$. With the circuit switched on, the change of $V_{bi}^{ph}$ should be correlated with $I_{ph}$. For simplicity, we assume the linear relation between $I$ and $\frac{dV_{bi}^{ph}}{dt}$ as

$$\frac{dV_{bi}^{ph}}{dt} = \frac{R_f}{Gt_e} I_{ph}, \tag{1}$$

where $R_f$ is the recombination factor that may be correlative to the specific distribution of PGHs in ISCR. To simplify the problem, we assume that $R_f$ is a constant, which is sufficient to explain our experimental results.

After optical charging, the photocurrent can be expressed as

$$I_{ph} = I_0 \exp(-\frac{q_e V_{bi}^{ph}}{kT})[\exp(\frac{q_e V}{kT}) - 1], \tag{2}$$

where $V$ is the bias voltage, $q_e$ is the electric charge of an electron and $k$ is the Boltzmann constant. In our experiments, $V$ always increases at a constant rate ($v_r$), where $V = v_r t$. From equations (1) and (2), a differential equation for the CPC effect can be formed as

$$\frac{kT}{q_e} \cdot \frac{dI_{ph}}{dt} = \frac{v_r I}{1 - \exp(-\frac{q_e v_r t}{kT})} - \frac{R_f}{Gt_e} I_{ph}^2 \quad (3)$$

Considering the initial conditions, we obtain the $I$-$V$ relation as

$$I_{ph}(V, V_{bi,0}^{ph}) = \frac{I_0 v_r Gt_e [\exp(\frac{q_e V}{kT}) - 1]}{I_0 R_f [\exp(\frac{q_e V}{kT}) - \frac{q_e V}{kT}] + v_r Gt_e \exp(\frac{q_e V_{bi,0}^{ph}}{kT})} \quad (4)$$

where $V_{bi,0}^{ph}$ is the initial $V_{bi}^{ph}$. Interestingly, this equation predicts the existence of a saturation current ($I_s = Gt_0 v_r/R_f$) at a large bias voltage. In the experiments, we decrease $v_r$ to avoid junction breakage, and increase the upper limit of the bias voltage. Fig. 5b shows the observed saturation behavior and its fit curve. The fit parameters used to generate the fit curve are: $I_0 = 1.3$ nA, $V_{bi,0}^{ph} = 9.8$ V and $R_f/G = 30.8$ Ω. It is worth pointing out that $R_f/G$ has the same dimension as resistance, indicating their similar physical connotation. As we have known, resistance is applied to estimate the ability to generate current by transporting carriers. As an analogy, $R_f/G$ may reflect the ability to generate photocurrent by releasing the photocarriers stored in ISCR. The theoretical simulation is highly consistent with the experimental data, supporting our explanation in relation to the CPC effect. Furthermore, this type of saturation behavior can also be understood qualitatively. $v_r$ signifies the ability to activate the trapped photocarriers, whereas $R_f$ indicates the tendency to consume them. Because of the opposition between the above processes, the saturation current can be achieved while reaching a dynamic balance.

From equations (1) and (4), we have

$$V_{bi}^{ph} = V_{bi,0}^{ph} + \frac{kT}{q_e} \ln \frac{I_0 R_f [\exp(\frac{q_e V}{kT}) - \frac{q_e V}{kT}] + v_r G t_e \exp(\frac{q_e V_{bi,0}^{ph}}{kT})}{I_0 R_f + v_r G t_e \exp(\frac{q_e V_{bi,0}^{ph}}{kT})} \quad . \quad (5)$$

This equation shows $V_{bi}^{Ph}$ as a function of $V$. It may be noted that $V_{bi}^{ph}$ is almost linear to a large V (see Supplementary Note 9). The inset in Fig. 5b exhibits the theoretical and experimental $V_{bi}^{ph}$-$V$ relation. The method to obtain $V_{bi}^{ph}$ is exhibited in Supplementary Note 10. The fit parameters used to generate the fit curve are: $I_0$ = 1.3 nA, $V_{bi,0}^{Ph}$ = 1.2 V and $R_f/G$ = 6.2 Ω. The good agreement between them implies that the assumption, on which equation (1) is based, is reasonable. In addition, the above model cannot predict the discharge process under negative bias voltage. The reverse bias makes the recombination of electron-hole take place outside ISCR, which may cause the inapplicability of the model (see Supplementary Note 1).

As the first CPC device, the WSe$_2$/Q2DEG heterostructure shows the potential applications for optical memory, optical information, self-excited devices and energy storage. One crucial advantage of the CPC effect is the ability to achieve the conversion and storage of optical energy simultaneously. To the best of our knowledge, no photo-electric devices that implement such functionality have been developed prior to this work. The process of optical charging is completely self-excited, with no need of electricity. This feature may be very useful for photo detection especially in an imaging system. By applying CPC photosensors in photo-detection circuits of charge-coupled device or complementary metal-oxide-semiconductor system,[38,39] the optical images can be directly converted into electric signals without any electric driving, and be long-term recorded until an erasing operation, which implies a new imaging device with very low power consumption and writable optical memory. Moreover, due to the infinite lifetime of the photocarriers stored in the junction, randomly-scattered photons in the environment may be gathered and harvested gradually so as to generate a large photocurrent, which is an unusual occurrence with entropy loss. The ability to accumulate photocarriers in CPC photosensors allows the achievement of high-quality imaging by long-time exposure, especially in an environment with very weak light.

For this new field, the infinite-lifetime photocarrier generation mechanism is a key factor causing the CPC effect. Similar to superconductivity, the CPC phase tends to eliminate the thermal loss inside the device, where $T_c$ is the temperature of the phase transition point. Our model has clearly explained the electric transport behaviors of infinite-lifetime photocarriers, but their generation mechanism requires further research. In the future, more attempts will be made to raise the $T_c$ to room temperature and enhance system performance. Other *p*-type 2D materials may be used to form p-n junctions with Q2DEG by replacing $WSe_2$. Modification of the CPC effect could also be considered via the insertion of an insulating layer between *p*- and *n*-type materials. Additionally, the application of other techniques, including electrical gating and chemical doping, is necessary to open more possibilities and reveal further useful properties.

In summary, $WSe_2$/Q2DEG heterostructure–based CPC devices show promise for high-efficiency photoelectric conversion and energy storage. Our experiments demonstrate that the infinite-lifetime photocarriers and insulating *p-n* interface are two key factors in determining the CPC effect. Further, the photocarriers can be completely self-excited and gradually accumulated without a thermal loss, and we have built a theoretical model to explain this striking effect. Our work opens up new opportunities to achieve high-performance collection of optical information with multiple benefits for future memory and energy applications.

**Methods**

**Fabrication of WSe2/Q2DEG heterostructures.** WSe$_2$ flakes were mechanically exfoliated from a piece of bulk single crystal onto STO substrates. A detailed description of the fabrication process has been outlined in Supplementary Fig. 1. Through a lithographic technique, half of a WSe$_2$ flake was coated by photoresist, and the other half was exposed to air. Then, the sample was irradiated for 14 minutes by an Ar$^+$ ion beam with a beam voltage of 200 V and beam current of 5 mA. A water-cooled sample holder was used during the etching process.

**Characterization.** Few-layered WSe$_2$ flakes are first identified by an optical microscope. Atomic force microscopy was used to determine the thickness of the samples.

**Photoelectric measurement.** I-V measurements were performed by the standard two-terminal configuration with Keithley 2400 and 6517B Source Meters. Photoelectric responses were measured in a physical property measurement system (Quantum Design). All electrical measurements were performed in an He atmosphere. Power-tunable light, supplied by semiconductor lasers with wavelengths of 405 ~ 808 nm, was transmitted into the sample chamber through an optical fiber bundle.

**Acknowledgements**

This work has been supported by the National Natural Science Foundation of China (Grant No. 11504254, 11304089, 51802210, 61875145, 11704272, 11574227, 11974304, 11734016). This work is also supported by Suzhou Key Laboratory for Low

Dimensional Optoelectronic Materials and Devices (SZS201611), Jiangsu Key Disciplines of the Thirteenth Five-Year Plan (20168765), the Natural Science Foundation of the Jiangsu Higher Education Institutions of China (18KJB430023), Natural Science Foundation of Jiangsu Province (BK20180970) and Fund of Shanghai Natural Science Foundation (grant no. 18ZR1445800).

**Author contributions**

Y.C.J. conceived the project, designed the experiments and wrote the manuscript. Y.C.J. developed the fabrication methods with input from A.P.H. and G.Z.L. A.P.H., Y.C., X.T.G. and Y.C.J. performed all the measurements. Y. C., A.P.H., R.Z. and Y.C.J. made all the figures. The theoretical model was developed by Y.C.J. with input from H.L. and R.Z.. M.S.L. and W.D.H. prepared thin $WSe_2$ flakes on the STO substrates. Y.C.J., A.P.H., G.Z.L. J.Q.N. and R.Z. analyzed the data, and Q.Y.W., J.G. and W.D.H. commented on the manuscript. All authors contributed to the discussion on the experimental results.

**Data availability**

All relevant data is available from the authors.

**Code availability**

All relevant code not already in the open literature can be requested from the authors.

# Competing financial interests

The authors declare no competing financial interests.

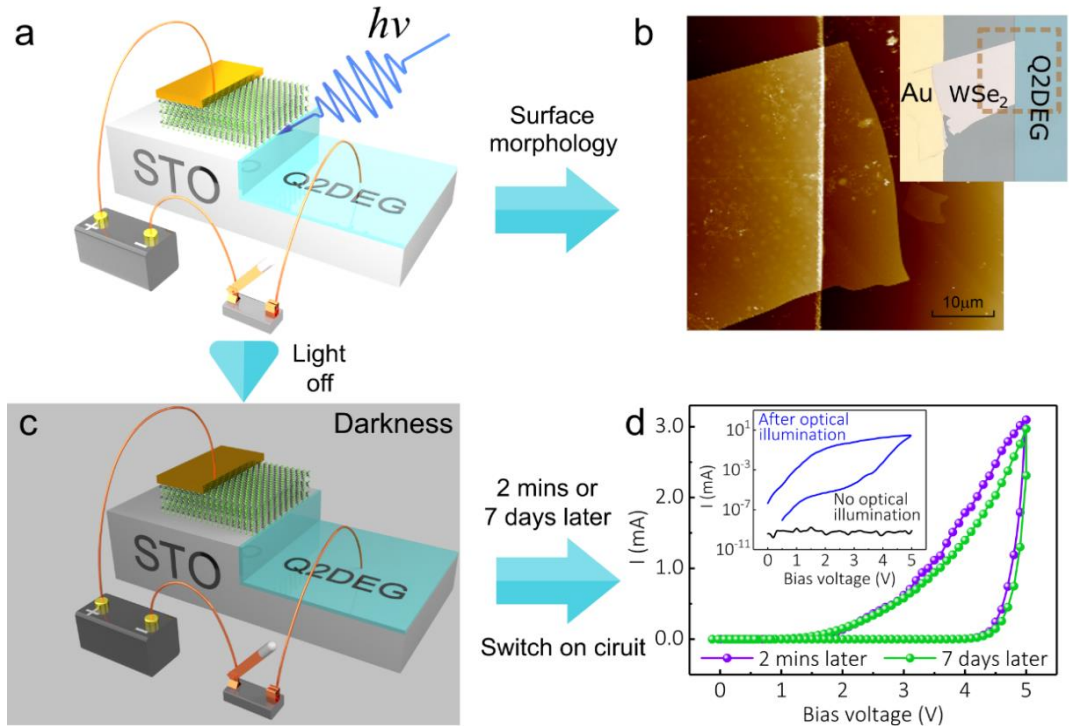

**Figure 1 | Device and photoelectric response. a** Schematic of WSe$_2$/Q2EG heterostructure under 405-nm, 16 mW/cm$^2$ optical illumination. Exposure time ($t_e$) is 6 s and circuit is cut off. **b** AFM image showing the surface morphology. Inset: photograph of the device. **c** Device placed in darkness for 2 mins or 7 days at 30 K. **d** I-V loops measured in darkness 2 mins later and 7 days later, respectively. Inset: I-V curves without and after optical illumination.

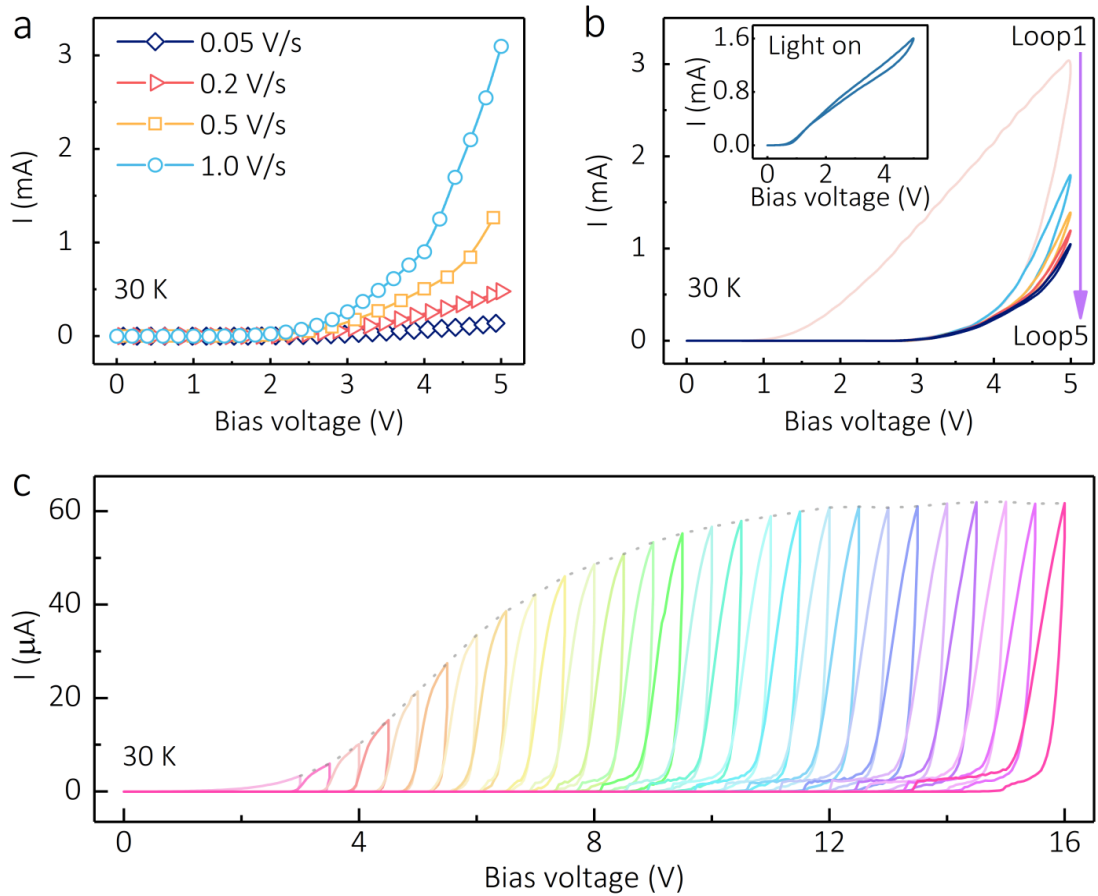

**Figure 2 | CPC effect. a** Dependence of photocurrent on $v_r$ after full charging. For simplicity, only forward photocurrent is shown. **b** I-V loops in multiple circles after a single optical charge, showing the fading photocurrent as the loop number increases. Inset: I-V loop under continuous optical illumination. **c** I-V loops following several voltage ranges of 0 – 4 V, 0 – 4.5 V, 0 – 5 V, 0 – 5.5 V…… 0 – 16 V in turn after a single full charging, with $v_r$ = 0.1 V/s.

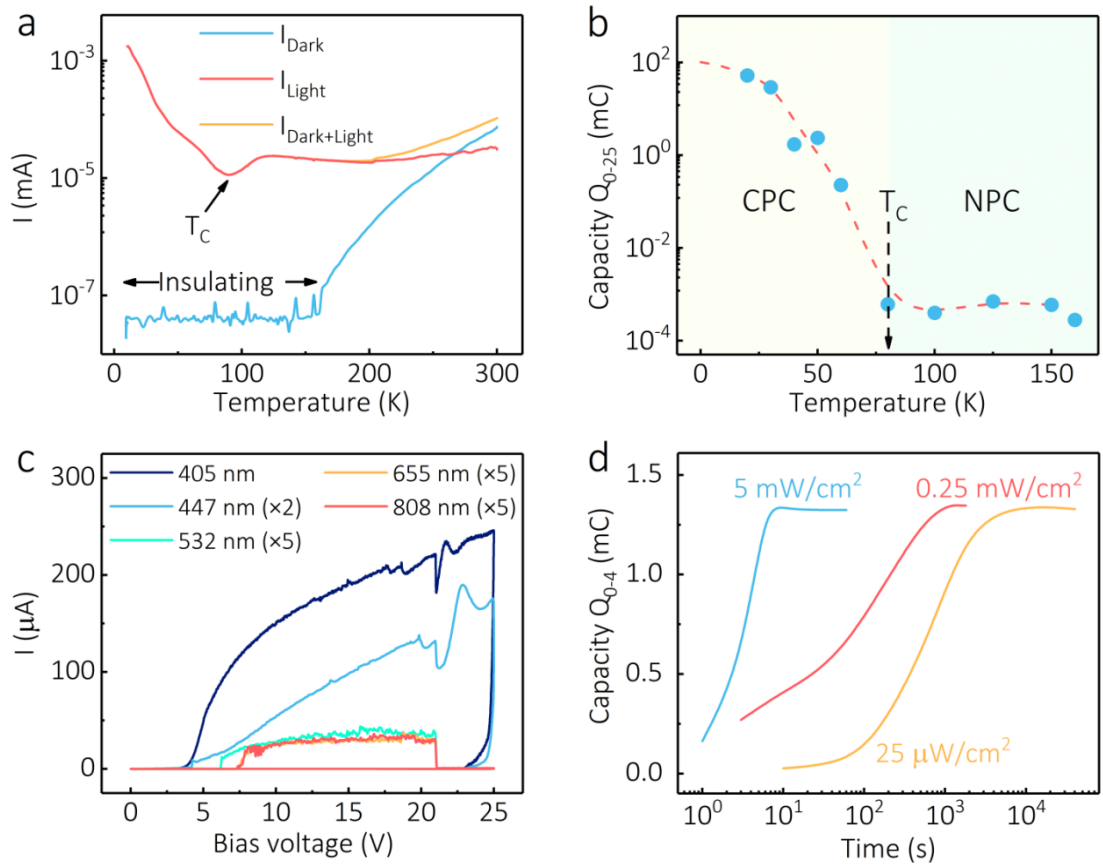

**Figure 3 | Temperature dependence and charging process of CPC effect. a** Temperature dependence of dark current, photocurrent and overall current under a 1 mW/cm$^2$ optical illumination at 4 V. **b** Charge quantity of stored photocarriers as a function of temperature. **c** I-V loops formed by using lasers of different wavelengths to charge the device. All measurements are performed with full charge and $v_r = 0.25$ V/s. Photocurrent at 447 nm is amplified twofold, and those at 532 nm, 655 nm and 808 nm is amplified fivefold, respectively. **d** Charge quantity of stored photocarriers depending on exposure time with different power densities of 405-nm light, showing the same saturation charge quantity.

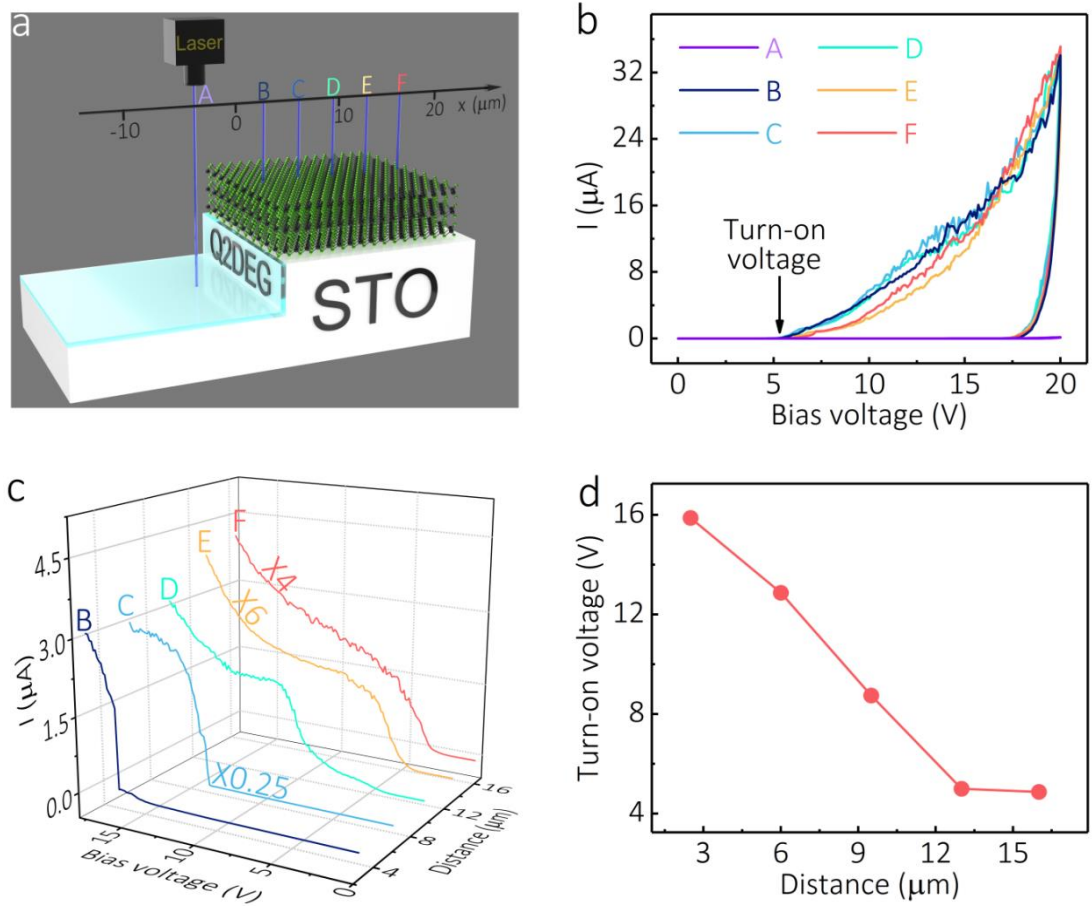

**Figure 4 | Generation and storage of photocarriers. a** Schematic showing that 405-nm MLB of 0.7-μm radius illuminates different positions on the WSe$_2$/Q2DEG heterojunction. **b** I-V loops after several optical charges with MLB illuminating the positions of A to F, where $v_r$ = 0.5 V/s, $P_{MLB}$ = 5 μW, $t_e$ = 40 s and $T$ = 30 K. **c** I-V characteristics measured from 20 V to 0 V, with continuous MLB illuminating the positions of B to F. Photocurrents at C, E and F are multiplied by 0.25, 6 and 4, respectively. **d** Turn-on voltages as a function of the distance between MLB and the interface. The data is obtained from **c**.

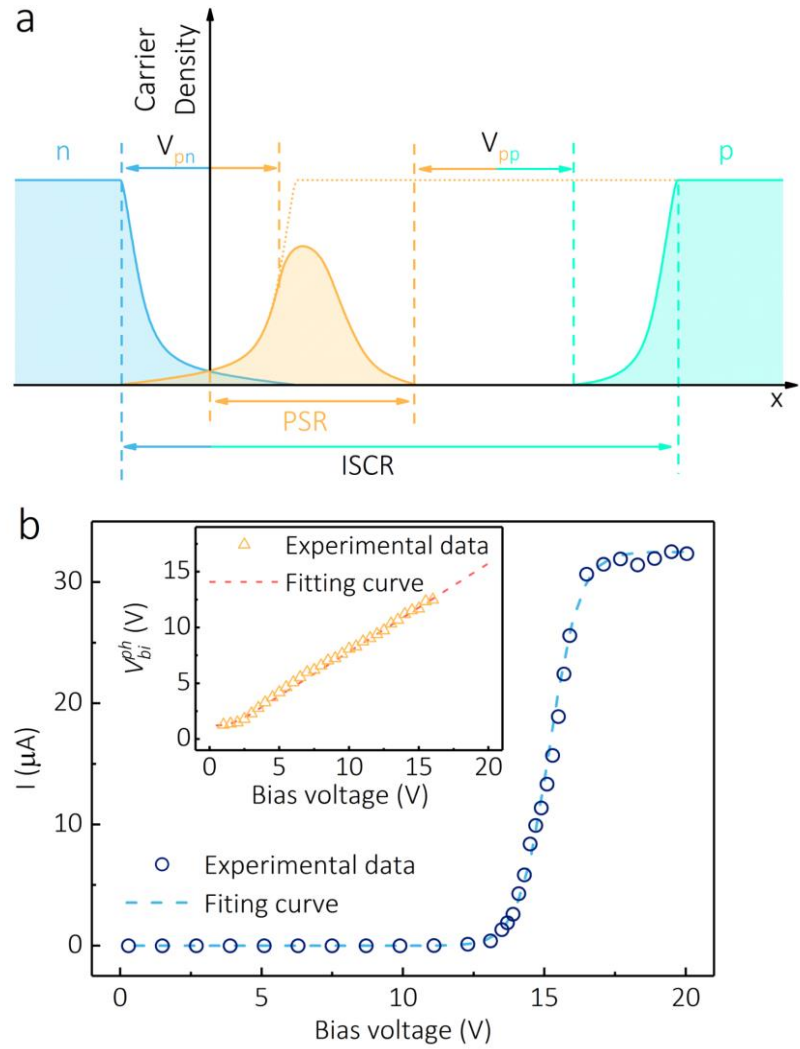

**Figure 5 | CPC mechanism. a** Schematic showing the carrier distribution after an illumination of light pulse. Infinite-lifetime PGHs are stored in the ISCR. **b** Fitting curves showing the discharge photocurrent with bias applied, where $v_r$ = 0.001 V/s, $P_{power}$ = 1 mW/cm$^2$ and $t_e$ = 1 s. Inset: the linear relation between $V_{bi}^{P,h}$ and $V$ after full optical charging, where $v_r$ = 0.1 V/s, $P_{power}$ = 5 mW/cm$^2$ and $t_e$ = 6 s.